\def\-{~-~}
\def\+{~+~}
\def\={~=~}
\def\eq{~\equiv~}
\def\tilde{\widetilde}
\def\beq{\begin{equation}}
\def\eeq{\end{equation}}
\def\beqn{\begin{eqnarray}}
\def\eeqn{\end{eqnarray}}
\def\bb{\begin{eqnarray*}}
\def\ee{\end{eqnarray*}}
\newcommand{\calle}[1]{(\ref{#1})}
\def\S{Schr\"odinger\  }
\def\Vy{\tilde V(y)}
\def\Psiy{\tilde \Psi}
\def\pseudoH{\tilde{\bf H}}
\def\Y{\tilde Y}
\documentstyle[preprint,revtex,eqsecnum]{aps}
\begin{document}
\draft
\preprint{CLNS-93/1188,HWS-9313}
\medskip

\begin{title}
\bf{Separation of Variables and Exactly Soluble\\
    Time-Dependent Potentials in Quantum Mechanics}
\end{title}

\author{\bf{Costas John Efthimiou}$^\dagger$ and \bf{Donald
Spector}$^\ddagger$}

\begin{instit}
$\dagger$ {Newman Laboratory of Nuclear Studies}\\
Cornell University\\
{Ithaca, NY 14853, USA}
\end{instit}

\vskip .2cm

\begin{instit}
$\ddagger${Department of Physics, Eaton Hall}\\
{Hobart and William Smith Colleges}\\
{Geneva, NY 14456, USA}
\end{instit}

\begin{abstract}
We use separation of variables as a tool to identify and to analyze exactly
soluble time-dependent quantum mechanical potentials.  By considering the most
general possible time-dependent re-definition of the spatial coordinate, as
well as general transformations on the wavefunctions, we show that separation
of variables applies and exact solubility occurs only in a very restricted
class of time-dependent models.  We consider the formal
structure underlying our findings, and the relationship between our results
and other work on time-dependent potentials.
As an application of our methods, we apply our
results to the calculations of propagators.
  \end{abstract}
\vskip 2mm
\hrule
$\dagger$ {costas@beauty.tn.cornell.edu}\\
\indent$\ddagger${spector@hws.bitnet}
\newpage

\section{Introduction}

In a recent paper, quantum mechanical systems with explicit time
dependence were studied via separation of variables \cite{RogersSpector}.
In that paper,
the authors identified changes of variables that allowed certain
time-dependent quantum mechanical systems to be treated using
separation of variables. (For convenience, we will use the term
{\it separable} to describe the situation in which separation of
variables can be applied effectively.)
The applicability of separation of
variables in these models means that for these time-dependent systems,
there exists a mathematical structure and formalism analogous to that
of the usual time-independent Schr\" odinger equation. As an
application of their approach, the authors used separation of variables, in
combination with both operator and power series methods, to solve
exactly the harmonic oscillator with inverse frequency linear in time.

That paper left open some important questions. First, can the
technique of \cite{RogersSpector} be used to identify
exactly soluble time-dependent
generalizations of other exactly soluble time-independent systems? Second,
how special was the change of variables that paper used to find
separable models? Would other
changes of variables have yielded other models? Third, can we use this method
to simplify or make possible other exact calculations in time-dependent
quantum mechanical theories?  And finally, what connections can be drawn
between the use of separation of variables and other approaches to
time-dependent quantum mechanical systems?
This paper addresses these questions.

Our central findings are as follows.  First, we show explicitly
how to construct exactly soluble time-dependent generalizations of any exactly
soluble time-independent model. We do this using a time-dependent
redefinition of the spatial coordinate which is linear in the spatial
variable. Second, we show that this linear change of variables, although
apparently quite special, is actually quite general: if we consider
{\it any} time-dependent re-definition of the spatial variable (but leave the
time coordinate as the second variable), exactly the same models are
uncovered as by the simple linear transformation. Including arbitrary
multiplicative transformations of the wavefunctions,
we are able to obtain a larger family of
separable and exactly soluble models. We find that separability singles
out a unique possible wavefunction transformation, and makes possible a
larger (but still specified) set of useful changes of variables.
Third, as an
application of our method, we present exact calculations of propagators in
exactly soluble time-dependent models, combining our techniques with those of
shape invariance.  And, finally, at various points in the paper,
we comment on the connections between our
work and the existing literature on time-dependent systems, with a focus
in one section on establishing the quantum generalization of the work
of \cite{LewisLeach} on classical Hamiltonians with an invariant
quadratic in momentum, and showing that this leads to exactly the same
set of models as separability does.

One of the key results of this paper is that we reduce the determination
of whether a time-dependent potential is exactly soluble to the
determination of whether
an associated time-independent potential is exactly soluble.
We do this through the use of separation of variables as the central
aspect of our approach.
As is well-known, this method
reduces a multi-dimensional problem to two or more lower dimensional ones. The
method of separation of variables is very powerful, and so has been applied
broadly. As an excellent recent example, Sklyanin has introduced the so-called
functional Bethe ansatz as an alternative to the algebraic Bethe ansatz, in
order to remove the restrictions inherent in the second approach. This new
method is a hybrid of the quantum inverse scattering method and the
method of separation of variables \cite{Sklyanin}.

The existing
literature on time-dependent quantum mechanics already
covers a variety of other
approaches. A sampling of such articles includes
\cite{Ray,Dodonovetal,Kaushal,Grosche}, as well as the
references therein. These articles provide a variety of approaches to
time-dependent quantum mechanics, for the most part with a focus on exact
solubility from the start. In this way our work is somewhat different; we
take separation of variables as our central criterion, and study its
implications.
Our interest in separation of variables is twofold:
separability implies the existence of a mathematical structure in a theory
analogous to the structure associated with the ordinary time-independent \S
equations; and, as we see from our work, particularly as it stands in
relation to the work in \cite{Ray,Dodonovetal,Kaushal,Grosche}
 (e.g., our methods identify all the
exactly soluble models found with other approaches in those references),
separation of variables also appears to be
an appropriate stepping stone in at least one path toward uncovering
and classifying the exactly soluble time-dependent models.
The agreement among these various papers (including ours) where they
overlap suggests the existence of some fundamental principles
that connect their various approaches.
We will comment on the connections between our work and that
of these other papers where relevant.

For the sake of simplicity, we will restrict our attention in
this paper to systems with only one spatial dimension, but it
will be readily apparent that our techniques can be applied without
modification to
higher dimensional models as well (see also \cite{Kaushal}).

Overall, we feel time-dependent quantum mechanics is still
inadequately understood.  For example, an explicitly time-dependent property
analogous to shape invariance --- which has greatly enriched our
understanding of exact solubility in the time-independent case --- is still
elusive. We believe the restrictive nature of
the results we have obtained using
changes of variables with separation of variables is significant,
because this helps us to determine what the actual range of exactly
soluble time-dependent models is; this can serve to focus the search
for a fundamental explanation of exact solubility in the time-dependent case.

\section{Exactly Soluble Time-Dependent Potentials}

To begin, we consider non-relativistic quantum mechanics in the \S
picture.
The \S equation in position space reads
\beq
   {\bf H}\Psi(x,t)
   \= i\hbar {\partial \Psi(x,t) \over \partial t}~,
\label{schrodinger}
\eeq
where the Hamiltonian, with time-dependent potential, is
\beq
      {\bf H} = -{\hbar^2 \over 2m}{\partial^2 \over \partial x^2} +
             V(x,t)~.
\label{hamiltonian}
\eeq
Our goal is to identify time-dependent potentials $V(x,t)$ such that
the \S equation \calle{schrodinger} is exactly soluble.
Obviously, separation of variables in terms of $x$ and $t$ will not
be helpful in this case, so we attempt to find a change of
variables for which separation of variables will be useful. To this
end, we define a new variable $y = f(t)x$.
In order that this change of variables be productive, we require that the
potential factorizes in terms of these new coordinates, that is, we require
that
\beq
      V(x(y,t),t)\= g(t)\, \Vy + h(t)~.
\label{separationofpotential}
\eeq
The time-dependent
\S equation now reads
\beq
    {-{\hbar^2 \over 2m}f^2(t) {\partial^2 \over \partial y^2}\Psiy(y,t)
    + \Bigl({g(t) }\,\Vy+h(t)\Bigr)\,\Psiy(y,t) = i\hbar{\dot f(t) \over
    f(t)}y{\partial \over \partial y}\Psiy(y,t) + i \hbar {\partial
    \Psiy(y,t) \over \partial t}~.}
\label{newschrodinger}
\eeq
We use $\Psiy$ to denote the wavefunction as a function of $y$ and $t$; thus
$\Psi(x,t)=\Psiy(y(x,t),t)$. Note that in
\calle{newschrodinger}, the partial derivatives with respect
to time are now taken
with $y$ fixed rather than with $x$ fixed.

Inspecting \calle{newschrodinger}, it is easy to see that
separation of variables
in $y$ and $t$ will work provided that
\beq
     {f(t)=(pt+q)^{-1/2}~,}
\label{fcondition}
\eeq
and that
\beq
   g(t) = a\, f^2(t)\=a\, (pt+q)^{-1}~,
\label{gcondition}
\eeq
where $a$, $p$, and $q$ are constants. The constant $a$ may be absorbed in
$\Vy$; thus, without loss of generality, we take $a=1$.
Hereon, we take the separation of variables conditions
\calle{separationofpotential}, \calle{fcondition},
and \calle{gcondition} to hold.
Now by  considering  solutions to the \S equation \calle{newschrodinger}
of the form $\Psiy(y)=T(t)Y(y)$, we see that the general solution to the \S
equation can be written as
\beq
    \Psiy(y,t)\=\sum_k\,c_k\, T_k(t)Y_k(y)~,
\label{generalsolution}
\eeq
where the $c_k$ are arbitrary constants, and where
 $Y_k$ and $T_k$
are the solutions of, respectively,
\beq
     -{\hbar^2\over 2m}\,{d^2Y_k \over dy^2}
      \+ i{\hbar p \over2}\,y{dY_k \over dy}
      \+  \Vy\, Y_k \= \gamma_k Y_k
\label{Yequation}
\eeq
and
\beq
     {i\hbar}\,(pt+q)\,{dT_k \over dt} - (pt+q)h(t) T_k
   \= \gamma_k \,T_k~,
\label{Tequation}
\eeq
with $\gamma_k$ a complex constant. {}From \calle{Yequation}, we see
that ${\rm Im} (\gamma_k) = -(\hbar p / 4)$.
The differential equation for $T_k$ can be solved exactly,
\beq
     T_k(t) =e^{-{i\over\hbar}\,\eta(t)}
             {\left(q + {pt}\right)}^{-(i\gamma_k / \hbar p)}~,
\label{Tsolution}
\eeq
where
\beq
\label{eta}
          \eta(t)\=\int_0^t\,dt\,h(t)~.
\eeq

Thus, for the above time-dependent Hamiltonians,
one recovers a mathematical structure formally analogous
to the one one finds for time-independent systems. Defining the
(non-Hermitian) pseudo-Hamiltonian
\beq
     \pseudoH \=
     -{\hbar^2\over 2m}\,{d^2 \over dy^2}
      \+ i{\hbar p \over2}\,y{d \over dy}
      \+  \Vy ~,
\label{pseudo}
\eeq
we see that the problem of solving the time-dependent
\S equation is equivalent to finding the eigenfunctions and (complex)
eigenvalues of $\pseudoH$.  (Note that in the limit that $p=0$, the
pseudo-Hamiltonian becomes the ordinary Hamiltonian, as it must.)

We are now in a position to determine the exactly soluble potentials as
follows.
Defining the wavefunction
\beq
      \Y_k(y)\=e^{-impy^2/4\hbar}\,Y_k(y)~,
\label{newY}
\eeq
we find that $\Y_k$ satisfies an {\it ordinary}
(time-independent) \S equation
\beq
\label{finalschrodinger}
     -{\hbar^2\over 2m}\,{d^2\Y_k \over dy^2}
      \+  {\tilde V}_-(y)\,\Y_k\=\epsilon_k\, \Y_k ~,
\eeq
where
\beq
\label{Vminus}
      {\tilde V}_-(y)\eq \Vy\-{p^2 m \over 8}\,y^2~,
\eeq
and where $\epsilon_k$ is a real constant given by
\beq
\label{energy}
     \epsilon_k\eq \gamma_k\+ i\,{\hbar p\over 4}.
\eeq
Thus, whenever ${\tilde V}_-(y)$ is an exactly soluble potential of ordinary
time-independent quantum mechanics, then the original time-dependent system is
itself exactly soluble.
Hence whenever
\beqn
       V(x,t)&\=& f^2(t)\,
       \left\lbrack
       {\tilde V}_-(f(t)x)\+{p^2m\over8}\, f^2(t)\,x^2
       \right\rbrack +h(t) \nonumber\\
\label{series1}
   &\=& {1\over pt+q}\left\lbrack {\tilde V}_-({x\over\sqrt{pt+q}})
            \+{p^2m\over8}\, {x^2\over pt+q}
       \right\rbrack +h(t)~,
\eeqn
and ${\tilde V}_-(y)$ is an exactly soluble potential
for the ordinary time-independent \S equation, the time-dependent
Hamiltonian of the \S equation \calle{hamiltonian} is exactly soluble.

In this way, we have identified a family of exactly
soluble time-dependent potentials to
generalize any exactly soluble time-independent potential.
Of course, much work has gone into the study of exactly soluble
time-independent potentials; most recent work has focused on
the uses of shape invariance \cite{Gendenstein}.
In fact,
all the well-known exactly soluble potentials of quantum mechanics can be
understood to derive their exact solubility ultimately from the property of
shape invariance. If a potential
${\tilde V}_-(y)$ is shape invariant, this means that
\beq
\label{VandW}
   {\tilde V}_-(y)\= W^2(y,\alpha)\- {\hbar\over \sqrt{2m}}\, W'(y,\alpha)~,
\eeq
and  $W(y,a)$ satisfies the condition
\beq
\label{shapeinvariance}
    W^2(y,\alpha)\+ {\hbar\over\sqrt{2m}}\, W'(y,\alpha)\=
    W^2(y,\sigma(\alpha))\- {\hbar\over\sqrt{2m}}\,
     W'(y,\sigma(\alpha))\+R(\alpha)~
\eeq
for some functions $\sigma(\alpha)$ and $R(\alpha)$.
Using the theory of supersymmetric quantum
mechanics \cite{LahiriBagghi}, the ``levels" $\gamma_k$ are given by
\begin{mathletters}
\beqn
\label{level1}
            \gamma_0&\=& -i{\hbar p\over 4}~,\\
\label{level2}
            \gamma_k&\=& -i{\hbar p\over 4}\+
                      \sum_{j=1}^{k-1}\, R(\alpha_j)~,~~~~~k\not= 0~,
\eeqn
\end{mathletters}
where $\alpha_j=\sigma(\alpha_{j-1})$ and $\alpha_0=\alpha$.

One of the intriguing features of the exactly soluble time-dependent
potentials  we have identified is
that, in addition to containing a term of essentially the same form as the
original potential, they all contain an additional time-dependent harmonic
oscillator term. We give a simple way to interpret this term below. But first
let us point out that because of the appearance of this term,
for the harmonic oscillator, and only for the
harmonic oscillator, will the time-dependent generalization simply involve
introducing a time dependence into the original coefficient of the
potential.  It is for this reason that \cite{RogersSpector} discovered no
exactly soluble systems other than the time-dependent harmonic oscillator.

As promised, we give here an instructive interpretation of the harmonic
oscillator term that appears in \calle{series1}. Using the definition of
${\tilde V}_-(y)$ in \calle{Vminus}, one  can write the \S
equation in $y$-space \calle{Yequation} in the following form:
\beq
\label{gaugeform}
    {1\over 2m}
    \left(  -i\hbar{d\over dy}
            -{pm\over 2}y      \right)^2 Y_k\+
     {\tilde V}_-(y)\, Y_k\=\epsilon_k\, Y_k ~.
\eeq
In other words, we can view \calle{Yequation} as describing
the coupling of a quantum
mechanical particle of mass $m$ to a background
electromagnetic field described by the
two-potential $(\varphi(y),A(y))=({\tilde V}_-(y),(pm/2)\,y)$, and
this necessitates the harmonic-oscillator--type term.
Invoking the gauge transformation $A(y)\rightarrow A(y)-\beta y$,
one can obtain
the eigenfunctions of the gauge-transformed Hamiltonian from those of the
original Hamiltonian by observing that under this transformation,
the eigenfunctions transform by
\beq
\label{phase}
     Y_k(y) \rightarrow
         \exp\Bigl({{i\over \hbar}{\beta^2\over 2}y}\Bigr)Y_k(y)~.
\eeq
In this way, then, the solutions to
the eigenfunction equation \calle{gaugeform}
can be obtained from those of the corresponding equation with an ordinary
kinetic term.  We point this out because we suspect that proper use of such
a gauge picture may yield a more fundamental way
to understand and to classify the
exactly soluble time-dependent models.

\section{Changes of Variables: The Potential}
The above set of exactly soluble models with
the potentials of \calle{series1} arises from a consideration of
the particularly simple change of variables $y=f(t)x$.  In order
to begin to develop a classification of separable and exactly
soluble time-dependent models, it is clearly essential to consider
other transformations, to see what separable and what exactly soluble
models these produce. We have done this, and the answer we find is
surprising.  Consider any change of variables
of the form $y=P(x,t)$, while keeping $t$ as the
second coordinate.  Then, neglecting Galilean transformations (as these are
essentially trivial),
we find that such a transformation can produce separation of
variables only when the function $P$ is in fact a function not of
$x$ and $t$ arbitrarily, but only of the ratio $x/\sqrt{pt+q}$, that is
$P(x/\sqrt{pt+q})$. (Of course, the potential must also separate
appropriately.)  Furthermore, the only exactly soluble models uncovered by
such transformations are exactly the exactly soluble models uncovered in
\calle{series1} by the very simplest transformation we considered.

We now offer a proof of this result.  The argument is rather technical,
and the reader interested primarily in our results might comfortably
omit this section on a first reading.

We will build towards the general proof
by first giving three preliminary results.
First, note that making a transformation $y=P(x/\sqrt{pt+q})$ for any
function $P$, will, given an appropriate potential, lead to separation of
variables in terms of $y$ and $t$.  Moreover, when the time-dependent
potentials are determined which are exactly soluble using separation of
variables and this transformation, one finds that they are exactly the same as
the potentials already uncovered in the previous section by the linear
transformation. The calculation proceeds exactly as in the preceding section;
in the interest of brevity, we will not present the details here, as they are
straightforward to obtain. (It is helpful in this calculation to invert the
transformation formally via $x = (\sqrt{pt+q}) P^{-1}(y)$.)

Second, consider a transformation slightly more general than the
linear transformation of the preceding section, namely the re-definition
of variables
\beq
\label{affine}
  y(x,t)=f(t)(x-u(t))~.
\eeq
Going through the same calculation as we did for the case $u(t)=0$, we find
that separation of variables is only possible for $\dot f \ne 0$ when, as
before,
\beq
\label{samef}
  f(t)={1\over\sqrt{pt+q}}~,
\eeq
and when
\beq
\label{showu}
  u(t) = a\sqrt{pt+q}+c ~.
\eeq
(When $\dot f = 0$, the only solution is a Galilean transformation; since
this is a trivial possibility, we ignore this case.) The additional
parameters $a$ and $c$ are easy to understand if we re-write \calle{affine}
using the results for $f(t)$ and $u(t)$, obtaining
\beq
\label{rewrite}
 y = f(t)(x-c)-a~.
\eeq
We see that the parameter $a$ corresponds to a shift in the origin of
$y$,  while $c$ corresponds to a shift in the origin of $x$. Thus these
parameters give us no new models. Any potential for which separation of
variables obtains when $a=0$ yields separation of variables for any value of
$a$, and so allowing the parameter $a$ adds no new models to our list
\calle{series1}. The parameter $c$ also gives us no new models.  If
separation of variables works for $V(x,t)$, it will not generally work for
$V(x-c,t)$ in terms of the same separating variables.  However, $V(x,t)$ and
$V(x-c,t)$ correspond to the same physical model.  Including the parameter
$c$ thus allows us to use separation of variables directly no matter what
origin one uses for $x$; the earlier transformation found all the same
models, but entailed choosing a particular origin for $x$.

The next result is a slight generalization of the preceding
one \calle{affine}. Consider now
a transformation of the form
\beq
\label{naffine}
  y = \varphi(t) (x-u(t))^n ~.
\eeq
Such a transformation can lead to separation of variables only when
$\varphi(t) = (pt+q)^{-n/2}$ (i.e., $f(t)^n$, where $f(t)$ is as above) and
$u(t)= a\sqrt{pt+q}+c$, as before. Rather than going through the same type of
calculation again, one can obtain this result from the preceding one. If
separation of variables holds in terms of $y$ and $t$, then it also holds in
terms of $z = y^{1/n}$ and $t$. In this way, we can reduce the problem of
finding useful transformations of the type in \calle{naffine} to finding
useful transformations of the type in \calle{affine}, which we already know.

We are now in a position to obtain our general result, namely that any
transformation of the form $y=P(x,t)$ leads to exactly the same
separable and
exactly  soluble models as we have already obtained. To begin, we will write
the  function $y = P(x,t)$ as a product,
\beq
\label{product}
  y=G(t)\prod_{j=1}^{k}\Bigl(x - u_j(t)\Bigr)^{n_j}~.
\eeq
For the purposes of our discussion, we take the $n_j$ to be positive but not
necessarily integral, although the restriction to positive $n_j$ is not
actually necessary. The $u_j$ must all be distinct. For
convenience, we define $N=\sum_{j=1}^k n_j$. We note that any $P$ which is
polynomial in the variable $x$ can be written in this  form. Since our proof
will hold for all polynomials of any order, it will also  hold in the limit
of an infinite but converging power series, which can be understood as the
limit of such polynomials. The set of transformations which we can handle
based on \calle{product} is thus a superset of those  functions which
have a power series in $x$.
In fact, the form of \calle{product} shows us that our argument
covers all functions except those with extreme singularities.

The essential idea behind our proof is to take the requirement that
separation of variables hold under this transformation, and look at the
implications of this requirement in different regions of space and time.
This will enable us to place restrictions on the possible forms of $G(t)$
and  the $u_j(t)$.

First, consider this transformation at large values of $x$, when $x>>u_j(t)$
for all $j$. In this region, the transformation is essentially
\beq
\label{largex}
  y \approx G(t)x^N ~.
\eeq
{}From our consideration of \calle{naffine}, we know that separation of
variables here requires that $G(t) \propto {(pt+q)}^{-N/2}$. Without loss
of  generality, we may re-scale the transformation so that
\beq
\label{Gis}
  G(t) = {1\over \sqrt{t+q}^N}~.
\eeq

Next we consider the case that $x$ is in the neighborhood of $u_j(t)$. Here
$x-u_j(t)$ is a small parameter. Expanding to lowest non-trivial order in
this parameter, we see that in this neighborhood
\beq
\label{nearuj}
  y \approx G_j(t) \Bigl(x-u_j(t)\Bigr)^{n_j}~,
\eeq
where $G_j(t)$ is
\beq
\label{Gjis}
  G_j(t) = G(t) \prod_{i=1, i\ne j}^k \Bigl(u_j(t)-u_i(t)\Bigr)^{n_i}~.
\eeq
In this neighborhood, the transformation takes the form of \calle{naffine},
and so we see immediately that we must be able to write each $u_j$ as
\beq
\label{ujis}
  u_j(t) = a_j \sqrt{t+q_j} + c_j~.
\eeq
Also, we see that $G_j$ is
\beq
\label{Gjnow}
  G_j(t) = \Bigl( {b_j \over \sqrt{t+q_j}}\Bigr)^{n_j}~.
\eeq
Thus
\beq
\label{constrainq}
  \prod_{i=1,i\ne j}^k \Bigl(u_j(t)-u_i(t)\Bigr)^{n_i}
        =\Bigl({b_j \over \sqrt{t+q_j}}\Bigr)^{n_j} \sqrt{t+q}^N~.
\eeq
Since we know the general form of the $u_j$'s, we know that the left side of
the preceding equation \calle{constrainq} has no divergences, and thus the
right side must have no divergences. This requires that $q_j = q$ in the
expression for $G_j$, which in turn dictates that $q_j=q$ in each of the
$u_j$, since, as we saw back in \calle{naffine}, the square root that appears
in $u(t)$ is the same square root as appears in $\varphi(t)$.

Putting all these results together, the equation \calle{constrainq}
now reads
\beq
\label{constrainc}
 \prod_{i=1,i\ne j}^k \Bigl((a_j-a_k)\sqrt{t+q~}+(c_j-c_k)\Bigr)^{n_i}
    = (b_j)^{n_j} \sqrt{t+q~}^{N-n_j}~.
\eeq
In order that the multiplicity of zeroes at $t=-q$ be the same on both sides
of the equation (and that there be no zeroes at any other value on the left
side, since there are none on the right side), we must have $c_j=c_k$ for all
values of $j$ and $k$, and $a_j\ne a_k$ for $j\ne k$. (This is consistent with
our requirement that all the $u_j$'s be distinct.) We define $c$ by
$c_j=c$ for all $j$.

We now see that in order for separation of variables to work, the
transformation in \calle{product} cannot be the most general product
possible, but must necessarily be of the form
\beq \label{ynow}
  y={1\over\sqrt{t+q}^N}
    \prod_{j=1}^k \bigl(x - a_j\sqrt{t+q} -c\bigr)^{n_j}~.
\eeq
We are free to redefine the origin along the $x$-axis; this does not change
the physical model. We therefore choose the origin such that $c=0$. Factoring
out the square roots (and recalling that $N=\sum_{j=1}^k n_j$), we see that
the most general possible transformation which has the possibility of
producing  separation of variables is of the form
\beq
\label{finaly}
  y=\prod_{j=1}^k \Bigl({x\over \sqrt{t+q}}-a_j\Bigr)^{n_j}~.
\eeq
Such a transformation is not a function of both $x$ and $t$ separately, but
rather is a function only of the combination $x/\sqrt{t+q}$. As
we discussed above, the separable and
exactly soluble models that are identified via separation of
variables following a  transformation $y=P(x/\sqrt{t+q})$ are exactly the
separable and exactly soluble models found using our original linear
transformation.

Thus by considering an arbitrary coordinate transformation which leaves time
as the second coordinate, we obtain no models that we did not already identify
by considering the simple linear transformation. This result is quite
restrictive, and it suggests that there is a rich mathematical structure
underlying both the applicability of
separation of variables to and the occurrence of
exact solubility in time-dependent quantum mechanics.
We would like to mention that we have tested our general proof by
examining a number of specific changes of variables.
In every case we have checked
(which includes a number of polynomial and rational functions for which
$y(x)$ can be inverted in closed form), we have verified that the only
changes of variables that can produce separability are the ones
identified by our general argument.

Thus we have shown that separation of variables only arises for
transformations $y=P(x,t)$ if $P$ is a function of $x/\sqrt{pt+q}$.
Let us re-emphasize here the twofold nature of the consequences of
this result: (1) there are no models that
admit separation of variables using an arbitrary transformation for which a
linear transformation does not suffice; (2) there are no exactly soluble
models that we find using separation of variables and an arbitrary coordinate
transform that we did not find using separation of variables and the linear
coordinate transform.
Thus, unless we consider changes of variables in which the time
variable is redefined (a possibility we do not explore in this paper),
we have, by using the simplest time-dependent
transformation, already identified
the most general exactly soluble time-dependent generalizations of the
exactly soluble time-independent models which are accessible by means of
separation of variables while using the original wavefunction.  In the next
section we include a
multiplicative transformation on the wavefunction along with changes
of variables, which enables us
to find a somewhat larger class of separable models.
Our results here (and in the next section)
demonstrate that the space of time-dependent exactly
soluble models is extremely constrained, and thus suggests that
separability may be a useful tool in classifying such models
and in identifying
the structure underlying their solubility.
\section{Changes of Variables: The Wavefunction}

The fact that the series of potentials found by the method described above
closely resembles the series of potentials discussed in \cite{Dodonovetal}
is very intriguing. It suggests that there should be a modification of the
above
prescription such that the potentials of \cite{Dodonovetal} --- and
perhaps more --- are exactly
reproduced. Here we present such a modification, which also serves
to help us ascertain the scope of models one can address
via separation of variables.
To this end, we define a new variable
\beq
\label{linear}
        y =  f(t)\,\Bigl(x\-\alpha(t)\Bigr)~,
\eeq
but also include now, in addition,
a transformation of the wavefunction
\beq
\label{ansatz}
              \tilde\Psi(y,t)\=\Xi(y,t)\,e^{\Phi(y,t)}~,
\eeq
where $\Psiy$  denotes the wavefunction as a function of $y$ and $t$, i.e.
$\Psi(x,t)=\Psiy(y(x,t),t)$.
(This ansatz is motivated
in part by an attempt to generalize the transformation in \calle{phase}.)
Without loss of generality we write the potential in the form
\beq
      V(x(y,t),t)\= g(t)\, \Vy + U(y,t)\+ g_0(t)~.
\label{potential}
\eeq
Furthermore, since we are interested in finding new models, we only
consider the case that $\Vy$ is not exclusively a quadratic
polynomial in $y$.

Using the above definitions, the \S
equation becomes
\beqn
     \-{\hbar^2\over 2m}\,{\partial^2\Xi \over \partial y^2}
     &\+&{\partial\Xi \over \partial y}\,
     \left\lbrack -{\hbar^2\over m}\,{\partial\Phi\over\partial y}-i\hbar
    {\dot f(t)\over f^3(t)} y +i\hbar{\dot\alpha\over f} \right\rbrack
    \nonumber \\
    &\+&\Xi\, \Bigg\lbrack -{\hbar^2\over 2m}\,{\partial^2\Phi\over\partial
y^2}
                          \-{\hbar^2\over 2m}\,\left({\partial\Phi\over
                           \partial y}\right)^2
       \+i\hbar\left({\dot \alpha\over f} -{\dot f(t)\over f^3(t)} y\,\right)
      {\partial\Phi\over\partial y}
    -{i\hbar\over f^2(t)} \, {\partial\Phi\over\partial t}
       \Bigg\rbrack \nonumber \\
      &\+& \Xi\, {1\over f^2(t)}\lbrack g(t) \tilde V(y)\, + U(y,t)\rbrack
       \= -{g_0(t)\over f^2(t)}\,\Xi\+
            {i\hbar\over f^2(t)} {\partial \Xi\over \partial t} ~.
\label{PHIschrodinger}
\eeqn
In order that
the coefficient of $\partial \Xi / \partial y$ be a function
only of $y$ (which is necessary for separability), we must have
\beq
\label{Phiis}
   \Phi(y,t)\=a(t)+b(t)\,{y^2\over 2}+c(t)\, y +  d(y)~,
\eeq
with
\beqn
\label{elimination1}
                    b(t)&\=& -{im\over\hbar}\,{\dot f\over f^3}~, \\
                    c(t)&\=& ~{im\over\hbar}\,{\dot\alpha\over f}~.
\label{elimination2}
\eeqn
Shifts of $b(t)$ and $c(t)$ by a constant from these values are allowed
consistent with separation of variables, but such constants can just
be absorbed into re-definitions of $a(t)$ and $d(y)$.

Since our goal in \calle{ansatz} is to make separation of variables
possible where it previously was not, we see that the choices
one makes for $a(t)$ and $d(y)$ are arbitrary, as these will
not change whether the
model is separable.  Hence, without loss of generality,
we make the convenient choices
$d(y)=0$ and
\beq
\label{afixing}
a(t) = {1\over 2}\ln f
   \+{i m \over 2\hbar}\int_0^t\,{ds\, {\dot\alpha}^2(s)}
   \-{i\over \hbar}\int_0^t\,{ds\, g_0(s)}~.
\eeq
Thus we see that given a transformation
to new variables \calle{linear}, there is
an essentially unique $\Phi(y,t)$
which has the potential to be useful in transforming the
wavefunction to obtain separability.
Using the value of $\Phi(y,t)$ we have now determined,
we see that the \S equation \calle{PHIschrodinger} simplifies to
\beqn
     \-{\hbar^2\over 2m}\,{\partial^2\Xi \over \partial y^2}
    &\+&\Xi\, \left\lbrack\, m{\ddot\alpha \over f^3}\, y\+
               {m \over2}{2\dot f^2 -f {\ddot f} \over  f^6}
              y^2 \,\right\rbrack \nonumber \\
      &\+& \Xi\, {1\over f^2(t)}\lbrack g(t) \tilde V(y)\, + U(y,t)\rbrack
       \=
            {i\hbar\over f^2(t)} {\partial \Xi\over \partial t} ~.
\label{NEWschr}
\eeqn

Since $\Vy$ is not simply a quadratic polynomial, in order for
separation of variables in $y$ and $t$ to work,
the function $g(t)$ must satisfy
\beq
   g(t) \=  f^2(t)~.
\label{Gcond}
\eeq

We now only have left to consider $f(t)$ and $\alpha(t)$.  Note first
that if the potential factorizes in the sense of
\calle{separationofpotential}, so that $U(y,t)=0$, separability
will only obtain provided that
\beq
\label{falphaU}
f(t) = {1\over \sqrt{\lambda t^2 + \mu t +\nu} }~,
\eeq
and with $\alpha$ fixed by requiring that the coefficient of the term
linear in $y$ in \calle{NEWschr} be independent of $t$.
This case makes contact with and generalizes the
results of the previous sections; more significantly, it reproduces
exactly the set of models investigated in \cite{Dodonovetal}.
So in the special case $U=0$, there are unique transformations
(which we have found)
of the spatial coordinate and the wavefunction
which can produce separation of variables.

In the more general case that $U\ne 0$, there is a somewhat larger
family of separable (and, in turn, a larger family
of exactly soluble) models.
Separability requires
that $U(y,t)$ exactly cancel the $y$ and $y^2$ terms, leading to
\beq
\label{Uhere}
     U(y,t)\=
    - m{\ddot\alpha\over f^3}\, y\-
               {m \over2}{2\dot f^2 -f {\ddot f} \over  f^3}
              y^2 ~.
\eeq
(We can shift $U(y,t)$ by any function of the form $U_0(y) f^2(t)$
or $\Gamma(t)$
consistent with separability, but these can just be absorbed into
redefinitions of $\Vy$ and $g_0(t)$, respectively.)

Now by  considering  solutions to the \S equation \calle{NEWschr}
of the form $\Xi(y,t)=T(t)Y(y)$, we see that, with $U(y,t)$ as
in \calle{Uhere}, the general solution to the \S
equation can be written as
\beq
    \Xi(y,t)\=\sum_k\,c_k\, T_k(t)Y_k(y)~,
\label{GENERALsol}
\eeq
where the $c_k$ are arbitrary constants, and where
 $Y_k$ and $T_k$
are the solutions of, respectively,
\beq
     -{\hbar^2\over 2m}\,{d^2Y_k \over dy^2}
      \+  \Vy\, Y_k \= \epsilon_k Y_k
\label{yEQUAT}
\eeq
and
\beq
     {i\hbar}\,{\dot T_k \over T_k} \=
   {\epsilon_k \,f^2(t)}~.
\label{tEQUAT}
\eeq
The differential equation for $T_k$ can be solved exactly,
\beq
     T_k(t) \=
           \, e^{-{i\over\hbar}\,\epsilon_k\tau(t)}
\label{tSOLUT}
\eeq
where
\beq
\label{etatau}
          \tau(t)\=\int_0^t\,dt\,f^2(t)~.
\eeq

Thus, whenever ${\tilde V}(y)$ is an exactly soluble potential of ordinary
time-independent quantum mechanics, then the original time-dependent system is
itself exactly soluble.
Grouping togehter several terms by defining
\beq
\label{gzero}
   h(t)\=
       -m\,\left(
                  {2\dot f^2 -f {\ddot f} \over  2f^2}\alpha^2
                 -\ddot\alpha\alpha  \right) +g_0(t)~,
\eeq
we see that whenever
\beqn
       V(x,t)\= f^2(t)\,
       {\tilde V}\Big(f(t)\Big\lbrack x-\alpha(t)\Big\rbrack\Big)\
    &\-& m\left( \ddot\alpha
          -\alpha {2\dot f^2 -f {\ddot f} \over  f^2}\right)\, x
        \nonumber\\
       &\-&{m\over2}\,  {2\dot f^2 -f {\ddot f} \over  f^2}
              x^2+h(t)~,
\label{series}
\eeqn
the model is separable, and that when, in addition,
${\tilde V}(y)$ is a standard exactly soluble potential
for the time-independent \S equation, the time-dependent
Hamiltonian of the \S equation \calle{hamiltonian} is exactly soluble.
In this way, we have identified a more general family of exactly
soluble time-dependent potentials.

\section{THE LEWIS--LEACH APPROACH}

The series of potentials \calle{series} that we obtained has been
found before, by Lewis and Leach
in a different but related context \cite{LewisLeach}.
They found that a classical Hamiltonian with explicit time
dependence has an invariant quadratic in the momentum if and only
if the potential is of the form \calle{series}.  (Their notation
is related to ours via the substitution $\rho(t) = 1/f(t)$.)
Lewis and Leach examined a purely classical problem, however; our
investigations are explicitly quantum mechanical.  In this section, we
show how to generalize \cite{LewisLeach} to the corresponding quantum
mechanical problem.
Remarkably, we find that a quantum mechanical Hamiltonian
will have an invariant quadratic in the momentum in exactly the same
cases that the classical Hamiltonian will, and that these are the
potentials we have identified by the condition of separability.

Any operator in quantum mechanics satisfies Heisenberg's equation
\beq
  {dI\over dt}\={\partial I\over \partial t}+{i\over\hbar}\,\lbrack
H,I\rbrack~.
\eeq
We are interested in the case for which there is an invariant $I$
which is quadratic in the momentum.
Because of the identity
\beq
\label{pAcom}
       \lbrack p, A(x) \rbrack = -i\hbar {\partial A(x)\over\partial x}~,
\eeq
we can always write $I$ in the form
\beq
\label{Iform}
     I\= f_2(x,t)\,p^2+f_1(x,t)\,p+f_0(x,t)~.
\eeq
Then one can easily derive that the functions $f_0(x,t)$, $f_1(x,t)$,
and $f_2(x,t)$
satisfy the following differential equations:
\beqn
      {\partial f_2\over \partial x}&\=& 0~,\\
      {\partial f_2\over \partial t}-{i\hbar\over2m}\,
      {\partial^2f_2\over\partial x^2}+{1\over m}\,{\partial f_1\over\partial
x}&\=& 0~,\\
      {\partial f_1\over \partial t}-{i\hbar\over2m}\,
      {\partial^2f_1\over\partial x^2}+{1\over m}\,{\partial f_0\over\partial
x}
       -2f_2{\partial V \over \partial x}&\=& 0~,\\
      {\partial f_0\over \partial t}-{i\hbar\over2m}\,
      {\partial^2f_0\over\partial x^2}+i\hbar\, f_2 {\partial^2 V\over\partial
x^2}
       -f_1\,{\partial V \over \partial x}&\=& 0~.
\eeqn
Notice that in the limit $\hbar\rightarrow 0$ we recover the classical
equations
of Lewis and Leach \cite{LewisLeach} as we expect. The above system
of differential equations for the quantum case can be solved in the same
manner as the classical equations.  Remarkably, the solution to the
system of quantum mechanical equations is virtually
identical to the solution to the
classical equations!  In particular, we see that $f_2$ is independent
of $x$, and that if we define ${\tilde f}_0(x,t)
=f_0(x,t) -i\hbar m {\dot f}_2(t)/2$, then ${\tilde f}_0$, $f_1$, and
$f_2$ satisfy exactly the classical Lewis-Leach equations.
We can use this to find easily that the form of all possible
time-dependent potentials that satisfy the quantum Lewis-Leach
equations is
given by \calle{series}, and the associated integral of motion is
\beq
\label{integral}
    I\={1\over 2}\,\left\lbrack\rho\,\left({p\over m}-\dot\alpha\right)-
    \dot\rho\,(x-\alpha)\right\rbrack^2
      \+\tilde V\left( {x-\alpha\over \rho}\right)
     ~,
\eeq
with $\rho(t) = 1/f(t)$, where $f(t)$ is the function from
the coordinate transformation \calle{linear}.
Note that this invariant is exactly the pseudo-Hamiltonian
defined earlier in this paper.

Of course, the existence of an integral of motion does not guarantee
exact solubility; rather, it appears as the analogue of separability. Exact
solubility will occur only for $\tilde V(y)$ of certain
functional forms. The shape invariance condition  we used earlier in
this paper, for example, does guarantee exact solubility of the
corresponding time-dependent model.
\section{Propagators}

The propagators for the potentials \calle{series} can be given, of course,
in terms of either the $x$ or the $y$ variables.  The respective
expressions are
\beq
\label{propaga1}
          K(x,x';t,t') = \sum_k\, \Psi_k(x,t)\, \Psi_k^*(x',t')
\eeq
and
\beq
\label{propaga2}
          \tilde K(y,y';t,t') = \sum_k\, \tilde\Psi_k(y,t)\,
            \tilde\Psi_k^*(y',t')
\eeq
are obviously related by
\beq
\label{transformation}
       K(x,x';t-t')\=\tilde K \left({x-\alpha(t)\over\rho(t)},
                        {x'-\alpha(t')\over\rho(0)};t-t'\right)~,
\eeq
since $y=(x-\alpha(t))/\rho(t)$.  (We find it useful to work in terms
of $\rho(t)=1/f(t)$ in this section.)
Using the wavefunction transformation \calle{ansatz}
and substituting in the propagator expression \calle{propaga2}, we find
\beqn
          \tilde K(y,y';t,t')&\=&
                       {\Big(\rho(t)\rho(t')\Big)}^{-1/2} \,
                      e^{\Phi(y,t)}\,e^{\Phi^*(y',t')}\,
            \sum_k \,
                      \Xi_k(y,t)\,\Xi^*_k(y',t') \nonumber\\
                  &\=&
                       {\Big(\rho(t)\rho(t')\Big)}^{-1/2} \,
                      e^{\Phi(y,t)}\,e^{\Phi^*(y',t')}\,
             \nonumber\\
           &&\qquad\times    \sum_k  Y_k(y)\,   Y_k^*(y') \,
           e^{-{i\over \hbar}\epsilon_k\,(\tau(t)-\tau(t'))}~.
\label{fullpropagator}
\eeqn
Noting that for the ordinary Schrodinger equation
\calle{yEQUAT} which arises for $\Xi(y,t)$, the propagator is
\beqn
           K_0(y,y';t,t')&\=& \sum_k \, Y_k(y)\,  Y_k^*(y') \,
           e^{-{i\over \hbar}\epsilon_k\,(t-t')}~,
\eeqn
we see that
\beqn
         \tilde  K(y,y';t,t')&\=&
                       {\Big(\rho(t)\rho(t')\Big)}^{-1/2} \,
                      e^{\Phi(y,t)}\,e^{\Phi^*(y',t')}\,
          K_0(y,y';\tau(t)-\tau(t'))~,
\label{k}
\eeqn
where $\tau(t)$ is as defined in \calle{etatau}.
Returning to the original variable $x$ by using
equation \calle{transformation},
and the results for $\Phi(y,t)$ contained
in \calle{elimination1}--\calle{afixing}, we find,
after some calculation, that
\beqn
\label{finalformula}
           K(x,x';t,t')\=
                       {\Big(\rho(t)\rho(t')\Big)}^{-1/2} \,
         &&{\rm exp}\left(-{i\over\hbar}\int_{t'}^t\,dt\,h(t)\,\right)\,
           {\rm exp}\left\lbrack-{im\over2\hbar}\int_{t'}^t\,dt\,
                 \left(\dot\alpha(t)-{\dot\rho(t)\over\rho(t)}\alpha(t)\right)
                  \,\right\rbrack\,\nonumber \\
          &\times &{\rm exp}\left\lbrack{{im\over2\hbar}
                  \left({\dot\rho(t)\over\rho(t)}x^2-
          {\dot\rho(t')\over\rho(t')}{x'}^2\right)}\right\rbrack\,
          \nonumber\\
 &\times &{\rm exp}\left\lbrack{{im\over\hbar}\left(\dot\alpha(t)-
 {\dot\rho{t}\over\rho(t)}
           \alpha(t)\right)x\-
           {im\over\hbar}\left(\dot\alpha(t')-{\dot\rho(t')\over\rho(t')}
           \alpha(t')\right)x'}\right\rbrack\nonumber\\
           &\times & K_0\left(
           {x-\alpha(t)\over\rho(t)},{x'-\alpha(t')\over\rho(t')};
            \tau(t)-\tau(t') \right)~.
\eeqn
Despite its length, this result is relatively elementary in
nature.  One should also be able to derive it in a straightforward
manner in the path integral formulation \cite{Grosche}, although
we have not attempted to do so.  One can easily check that this
propagator gives the correct result in the special
cases $\alpha(t)\not=0$, $\rho(t)=1$ and
$\alpha(t)=0$, $\rho(t)=\sqrt{rt^2+2pt+q}$, by comparison with
\cite{Dodonovetal} and \cite{Grosche}, where these special cases are
considered.

We now give several examples based on the general formula
\calle{finalformula}.  We will identify models in which our method
in combination with other techniques makes the calculation of
propagators relatively easy.  In some cases we will actually display
the propagators, while in other cases we will give the reader the
necessary ingredients, but, in the interests of space and clarity,
not present the full form of the propagator.

As a preliminary, we rewrite the potentials \calle{series}
in the form
\beq
\label{newform}
    V(x,t)\={1\over \rho^2(t)}\, \tilde V\left( {x-\alpha(t)\over \rho(t)}
            \right)
     \, -F(t)\,x\+{1\over 2}\,m\,\omega^2(t)\,x^2~,
\eeq
where the functions $\omega(t)$ and $F(t)$ are solutions to
\beqn
       \ddot\rho\+\omega^2(t)\,\rho&\=&0~,
\label{ar1} \\
       \ddot\alpha\+\omega^2(t)\,\alpha&\=&{1\over m}\,F(t)~.
\label{ar2}
\eeqn
We will use $\omega$ and $F$ in lieu of $\alpha$ and $\rho$ as convenient.

The first example we consider is the free particle
of time-independent quantum mechanics, i.e.  $\tilde V(y)=0$,
which is of course an exactly soluble model.
This implies that the forced oscillator with a time-dependent frequency
\beq
\label{oscillator}
     V(x,t)\= -F(t)\,x\+{1\over 2}m\omega^2(t)\,x^2~,
\eeq
is also an exactly soluble model.

We can arrive at the same model by starting with a simple harmonic oscillator
$\tilde V(y)=m\omega_0^2y^2/2$. Then by taking
$h(t)=-m\omega_0^2\alpha^2/2\rho^4$ we find exactly the potential
\calle{oscillator} with one minor modification. Now,  $\rho(t)$ is
not a solution of \calle{ar1}, but rather of
\beq
       \ddot\rho\+\omega^2(t)\,\rho\={\omega_0^2\over\rho^3}~.
\label{ar3}
\eeq
Equation \calle{ar2} remains unchanged. This model is still separable
and exactly soluble,
even though $\rho$ satisfies \calle{ar3}
rather than \calle{ar1}, as the effect of the extra term
in \calle{ar3}
can be absorbed into a re-definition of $\Vy$
\cite{footnote}.

It is known that the solutions
of equations \calle{ar1} and \calle{ar3} are related.
In particular, if $\rho_0(t)$ is a solution of equation \calle{ar3},
then $\rho_1(t)=\rho_0(t)\,{\rm sin}\omega_0\tau_0(t)$ and
$\rho_2(t)=\rho_0(t)\,{\rm cos}\omega_0\tau_0(t)$,
where $\tau_0(t)=\int_0^tds\,{1/\rho_0^2(s)}$, are solutions of equation
\calle{ar1} \cite{Lewis}.
Also, a particular solution for equation \calle{ar2} can be found
using the theory of Green's functions:
\beqn
\label{a1}
   \alpha(t)&\=&\int_{t_0}^t\,ds\,{F(s)\over m\omega_0}\, \rho_0(t)\rho_0(s)
    \,{\rm sin}\lbrack\omega_0(\tau_0(t)-\tau_0(s))\rbrack \nonumber\\
   &\=&\int_{t_0}^t\,ds\,{F(s)\over m\omega_0}\,
    \left\lbrack\,\rho_1(t)\rho_2(s)
    -\rho_1(s)\rho_2(t)\,\right\rbrack~.
\label{a2}
\eeqn
Alternatively, the above expression can be verified by direct substitution
in equation \calle{ar2}.

Thus,
given $\rho_0(t)$, we can use either the propagator of the free particle
 or the propagator of the
simple harmonic oscillator with constant frequency to obtain
the propagator of the potential \calle{oscillator}. Since the
resulting expression is rather long and complicated, we will not
write down the result here, as its evaluation is entirely
straightforward.

As a by-product, notice that the propagator of the time-independent
oscillator of quantum mechanics can be derived by the one of the free
particle as had been noticed in \cite{Efthimiou}.
In this case, $F(t)=0$,
$\omega(t)=\omega_0=constant$, and it is enough to take
$\alpha(t)=0$, $\rho(t)={\rm cos}\omega_0t$;
thus $\tau(t)=(1/\omega_0){\rm tan}(\omega_0t)$.
Using the propagator of the free particle
\beq
\label{freep}
    K_0(y,y';t,t')\=\sqrt{{m\over2\pi i\hbar(t'-t)}}\,
                 {\rm exp}\left\lbrack {im\over2\hbar}{(y-y')^2\over t-t'}
                          \right\rbrack~,
\eeq
we find the well-known expression
\beq
\label{hoti}
    K(x,x';t,t')\=
    \sqrt{{m\over2\pi i\hbar{\rm sin} \omega_0(t'-t)}}\,
                 {\rm exp}\left\{
              {im\omega_0\over2\hbar}\left\lbrack
            {(x^2+x'^2)
                {\rm cot}\omega_0(t-t')\-{2xx'\over{\rm cos}\omega_0(t-t')}}
                          \right\rbrack\right\}~.
\eeq

We now consider a related model, the free particle in
\beq
\label{example1}
     \tilde V(y)\=
                  \cases{ 0, &if  $~0<y<L_0~,$\cr
                          +\infty,&if $~ y\le0~~~{\rm or}
                                           ~~~y\ge L_0~.$\cr }
\eeq
The exact solubility of this model implies that the infinite
well with moving boundaries
\beq
\label{timewell}
      V(x,t)\=
                  \cases{ -F(t)\,x+{1\over2}\,m\,\omega^2(t)\,x^2~,
                   &if  $~\alpha(t)<x<\rho(t)\,L_0+\alpha(t)~,$\cr
                          +\infty,&if $~ x\le\alpha(t),
                                   ~x\ge \rho(t)\,L_0+\alpha(t)~,$\cr }
\eeq
is also exactly soluble. The propagator of the time-independent infinite well
can be written down effortlessly using the energy eigenvalues
$E_n=n^2\pi^2\hbar^2/2mL_0^2$ and eigenfunctions
$Y_n(y)=\sqrt{2/L_0}\,{\rm sin}({n\pi y/L_0)}$, giving
\beq
  K_0(y,y';t,t')\={2\over L_o}\,\sum_{n=1}^{+\infty}\,
  {\rm exp}\left(  -{i\over\hbar}\,
     {n^2\pi^2\hbar^2\over2mL_0^2}(t-t')\right)\,
  {\rm sin}{n\pi y\over L_0}\,
  {\rm sin}{n\pi y'\over L_0} ~.
\eeq
This can be written in terms of the Jacobi $\theta_3$-function,
\beq
\label{th3}
      \theta_3(u|z)\eq 1\+2\sum_{n=1}^{+\infty}\,
       e^{in^2\pi z}\,{\rm cos}(2nu)~,
\eeq
giving
\beq
\label{propag1}
K_0(y,y';t,t')\={1\over 2L_o}\,\left\lbrack
\theta_3\left( {\pi(y-y')\over2L_0}\Bigg|-{\pi\hbar(t-t')\over2mL_0^2}\right)
\-\theta_3\left( {\pi(y+y')\over2L_0}\Bigg|-{\pi\hbar(t-t')\over2mL_0^2}\right)
 \right\rbrack
\eeq
It is now an easy exercise again to write down the propagator for the
infinite well with moving boundaries in $(x,t)$-space.

Using the theory of supersymmetric quantum mechanics, Das and
Huang \cite{DasHuang} have developed a very elegant way of calculating
the propagators of the time-independent exactly soluble potentials
of quantum mechanics.  For example, using this method one can
obtain the exact propagator for
\beq
\label{example2}
     \tilde V(y)\= {\hbar^2\over 2m}\, {n(n+1)\over y^2}~,~~~~~n=1,2,3,\dots~,
\eeq
which is \cite{DasHuang}\cite{Khare}\cite{Efthimiou}
\beq
\label{propag2}
           K_0(y,y';t)\={{m\omega_0\sqrt{yy'}
                       \over\hbar {\rm sin}(\omega_0t)}}\,i^{-(n+{3\over2})}
           \, J_{n+1/2}\left({m\omega_0yy'\over\hbar{\rm
sin}(\omega_0t)}\right)
      \,e^{{i\over\hbar}{m\over2t}(y^2+y'^2)}~.
\eeq
Our results, then, show that not only is the time-dependent potential
\beq
\label{timepotential1}
       V(x,t)\=-F(t)\,x\+{1\over2}\,m\,\omega^2(t)\,x^2\+{\hbar^2\over 2m}\,
      {n(n+1)\over (x-\alpha(t))^2}~,
\eeq
exactly soluble, but also that its propagator can easily
be obtained from the propagator \calle{propag2} and our fundamental
propagator formula \calle{finalformula}.

As final example, we note that in \cite{DasHuang}, the propagator
is found for the exactly soluble potential
\beq
\label{example3}
             \Vy \={\alpha^2\hbar^2\over m}
           \,\left({-1\over {\rm cosh}^2(\alpha y)}\+{1\over 2}\right)~.
\eeq
Putting the results of this paper together with those of \cite{DasHuang},
then, we can easily find
the propagator for the potential
\beq
\label{timepotential2}
   V(x,t)\=
              - {\alpha^2\hbar^2\over m}\,
                   {1\over \rho^2(t)}\,{\rm cosh}^{-2}
       \left({x-\alpha(t)\over\rho(t)}\right)\-F(t)\,x\+
      {1\over2}\,m\,\omega^2(t)\,x^2~.
\eeq
The final expression for the propagator is lengthy, but simple to obtain.

Needless to say, one can in the same way find the propagators for
the time-dependent generalizations \calle{series}
of the other known exactly soluble
time-independent potentials of quantum mechanics.
The examples above have been
selected as a sampler of typical results and techniques.

\section{Conclusion}

In this paper, we have explored the applicability of separation of variables
to quantum mechanical systems with explicitly time-dependent potentials.
We have presented this work in one spatial dimension, but clearly the same
methods can be applied in higher dimensions as well. We
have found that only a very limited set of changes of variables and
transformations of the wavefunctions can produce
separability.  The uniqueness we find is remarkable, and it indicates
that the underlying principles that allow separation of
variables are quite restrictive.

Using as our central technique separation of variables, we have
also constructed
exactly soluble time-dependent generalizations of any exact soluble
time-independent model.  In fact, our method reduces the identification of
exactly soluble time-dependent models to the identification of exactly
soluble time-independent models.  This allows us to make contact with the
very powerful insights and techniques of shape invariance, and promises a
deeper understanding of exact solubility in the time-dependent case.

Our work also makes significant contact with other work on
time-dependent Hamiltonians.  For example, we have used
separation of variables to find the exactly soluble models
discussed by \cite{Dodonovetal}.  We have also worked out the
quantum generalization of \cite{LewisLeach}, and shown that the
existence of an invariant quadratic in the momentum occurs for
exactly the same time-dependent Hamiltonians as separability does.

Of course, we cannot claim to have identified or classified all exactly
soluble
time-dependent potentials. However,
the unique time-dependent generalizations
produced by {\it any} useful changes of variables suggests
that there is indeed a very
intricate structure underlying the exact solubility of these models.
To identify this structure --- ideally, to do so directly in ordinary position
space --- would be a significant contribution to our understanding of exact
solubility and to a possible classification of exactly soluble time-dependent
models.

Furthermore, the results of this paper coincide at various points with
the results of the references cited on
exactly soluble time-dependent quantum mechanics
 \cite{Ray,Dodonovetal,Kaushal,Grosche},
 despite the differing
approaches of all these papers (including ours).  This suggests that
separation of variables is a sound guide to finding exactly soluble
time-dependent potentials.  Given the variety of techniques that have been
applied to time-dependent quantum mechanics, all of them
of course producing mutually
consistent results, there is probably much progress to be made by
using these various approaches in tandem.

\acknowledgments
C.J.E. was supported in part by the National Science Foundation.
D.S. was supported in part by NSF Grant. No. PHY-9207859.




\begin{references}
\bibitem[*]{footnote} Using
such an alternative $\rho(t)$ will preserve separability for
any model in the list \calle{series},
but exact solubility will depend on the particular form of
the potential.
\end{references}
\end{document}